# CLAASIC: a Cortex-Inspired Hardware Accelerator

V. Puente, J.A. Gregorio
University of Cantabria
Santander, Spain
vpuente@unican.es

**Abstract—**

This work explores the feasibility of specialized hardware implementing the Cortical Learning Algorithm (CLA) in order to fully exploit its inherent advantages. This algorithm, which is inspired in the current understanding of the mammalian neo-cortex, is the basis of the Hierarchical Temporal Memory (HTM). In contrast to other machine learning (ML) approaches, the structure is not application dependent and relies on fully unsupervised continuous learning. We hypothesize that a hardware implementation will be able not only to extend the already practical uses of these ideas to broader scenarios but also to exploit the hardware-friendly CLA characteristics.

The architecture proposed will enable an unfeasible scalability for software solutions and will fully capitalize on one of the many CLA advantages: very low computational requirements and optimal storage utilization. Compared to a state-of-the-art CLA software implementation it could be possible to improve by 4 orders of magnitude in performance and up to 8 orders of magnitude in energy efficiency.

Embracing the problem's complex nature, we found that the most demanding issue, from a scalability standpoint, is the massive degree of connectivity required. We propose to use a packet-switched network to tackle this. The paper addresses the fundamental issues of such an approach, proposing solutions to achieve scalable solutions. We will analyze cost and performance when using well-known architecture techniques and tools. The results obtained suggest that even with CMOS technology, under constrained cost, it might be possible to implement a large-scale system. We found that the proposed solutions enable a saving of ~90% of the original communication costs running either synthetic or realistic workloads.

## 1 Introduction

Mammal brains have a distinct structure compared to other biological systems: the presence of the neo-cortex. The salient feature of this construction is that anatomically and functionally it is remarkably homogenous. Eighty years ago Lorente de Nó [32] discovered that the neo-cortex (from now on, cortex) is composed of distinguishable packs of neurons forming *columns*. Later, V. Mountcastle [35] anatomically detailed the structure of these columns, as approximately two millimeter high structures where a set of six layers can be distinguished. The columns are connected via low-range axons to other nearby columns (via Layer I) or other distant columns and the thalamus (via Layers V and VI). The structures are called cortical columns or micro-columns (See Figure 1.a). Nearby columns form larger packs, called cortical hyper-columns or macro columns (Figure 1.b) [12]. Regardless of the functionality of each zone, the cortex is highly regular. For example, the human cortex is a surface of ~1600 square centimeters and there are negligible anatomical differences throughout it [51]. This fact has puzzled neuroscientists for decades: how such a highly regular structure can be the underlying structure for the complex functional organization of the cortex [36].

In neuroscience the most accepted hypothesis[12] is that the cortex is some sort of memory system. The inputs are composed of incoming signals from the senses and the outputs are the actions on lower level (prewired) brain structures. After that, they are sent to the motor system and expressed as behavior. The main hypothesis is that the cortex is continuously building a model of the world according to the sensory information flow. This model is used to make predictions. The cortex scale, i.e. number of cortical columns and potential connectivity, simply determines the reach of these predictions.

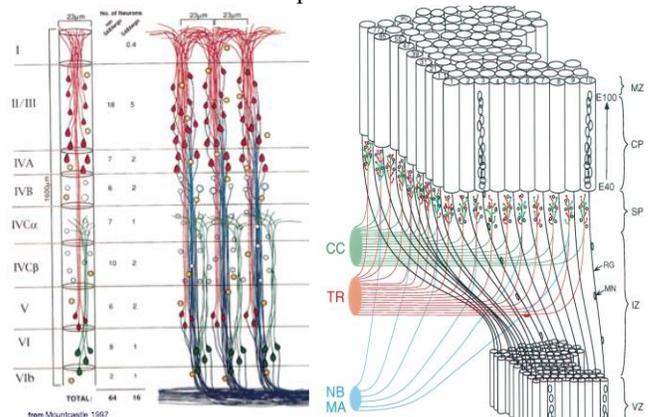

**Figure 1 (a) Cortical columns, (b) Cortical Macro-columns**

Initially George and Hawkins [21] hypothesized how the cortex might implement such a memory structure in order to fulfill the biological requirements. A theory has been built on the hypothesis that the cortex works as a self-associative[1] memory, hierarchically structured as a Hierarchical Temporal Memory (HTM) [25][34]. The theory, which is purely influenced by neuroscience observations, includes an algorithm, called the Cortical Learning Algorithm (CLA) that provides the rules for storing and retrieving information i.e. learning and making predictions. The idea has been used in practical problems such as anomaly detection, sequence prediction, pattern identification, natural language processing, etc. At this point in time, the theory can describe the behavior of the upper layers in the cortical column. Hopefully, the hierarchy's inner workings will also be understood soon. CLA[2,3] can roughly model the equivalent to a small group of hyper-columns, denoted as a *region*. There is intense work, making constant progress. Given this state, unarguably, other deep learning techniques (but highly specialized, and in most cases use off-line learning), such as [31], produce better results in known problems such as image classification. In contrast with these approaches, CLA has two remarkable properties: the core algorithm does not change from application to application, and, like in biological systems, it is continuously learning. Coincidentally, both properties are required to achieve Artificial General Intelligence (AGI), as defined by AI theorists [19]. In contrast to such highly specific techniques, which are supported by strong mathematical foundations, CLA is mainly influenced by experimental observations gathered by neuroscience. Even in this early stage, CLA has already proved its advantages over other state-of-the-art techniques, in anomaly detection [30], continuous unsupervised learning [17], natural language processing [53], etc.

---

[1] The current input sequence is used to address the forthcoming content (e.g. reciting the alphabet).

Currently, the progress made with CLA is based on software. There are many implementations, NuPIC [56] being the most remarkable one, which is supported by Numenta using OSS licensing (AGPLv3). Other companies, such as IBM [54], are working using their own implementation. Although the practical uses increase the complexity of such tools, the core algorithms are simple (but not simpler). This approach provides the necessary flexibility to explore new practical uses or new core algorithm variations. Nevertheless, the software limits the system size to a few thousands of columns, which might restrict the practical uses or advancements in hierarchy definition. To circumvent this problem, developing a feasible hardware seems necessary [54]. To our knowledge, the only effort in this direction is [8], which advocates the use neuromorphic hardware. Within the conventional ML realm, there are many specialized and limitedly flexible hardware implementations, [13][14][24].

While some one might argue that it could be too early to cast a specialized silicon product, given the algorithm and application development status, perhaps we need to start addressing the issues we might encounter later. In contrast with other machine learning (ML) approaches, such as Deep Neural Networks (DNN), CLA might bring some challenges that are not straightforward to tackle. Instead of precisely weighted connections and computing intensive matrix multiplications, CLA's foundation is a hyper-connected complex and highly dynamic topology to store and retrieve information. From a naïve hardware perspective, this is hard to achieve (a single column can potentially be connected to tens of thousands of different columns). Although emerging technologies, such as 3D stacking and Non-volatile memory might ease these stringent requirements, one might wonder if we can sketch a feasible implementation in a conventional CMOS process, as [13] successfully achieves.

Beyond that, CLA has a relevant advantage over DNN or other weight-based learning: it requires only a very low precision (around 4-bits might suffice) to add and compare instructions. Unfortunately, using conventional architectures software implementations cannot capitalize on this since topology handling requires frequent and costly data movement throughout the memory hierarchy, which might difficult the use of GPU programming models. A hardware implementation could greatly improve both the energy requirements and performance. It seems feasible to process many millions of samples per second within a constrained energy envelope. However, DNN, even using a significantly more costly multi GPU configuration, is far from achievable [6]. This will open up the possibilities of using CLA in problems that are hard to handle for conventional ML such as cognitive computing problems, where continuous unsupervised learning might be required. DNN has made progress in unsupervised learning [26]. Nevertheless, since this property is inherent to the CLA, it does not suffer the problems faced by DNN [7].

This paper explores this path, presenting the architecture of a feasible hardware implementation. In contrast with [8], we will use architectural methodologies/techniques similar to those used in commercially available products, such as general purpose chip multiprocessors. Again, looking at the biological properties of axons and dendrites, we define a system that uses a logical construct to fulfill the topological flexibility of CLA over an on-chip network. Given the low computational requirements, attaching some simple logic to the routers of this network and some memory to store the connectivity status, it could be possible to implement CLA without requiring complex and power hungry general-purpose CPUs or GPUs.

Consequently, like in biological systems, the network is the point towards which the system gravitates. We will focus our attention on the communication substrate and procedures to make CLA feasible. We will describe how, using a packet switched network and diverse techniques used in computer architecture, we can achieve a practical implementation. Different solutions will be presented to guarantee system scalability. We will analyze, through detailed simulation and using well-known modeling tools, the temporal and energy requirements of the system. The set of proposals introduced minimizes communication overheads. The combination of all these techniques on average reduces the network delay and active energy requirements by ~90%. Since communication seems to be the most demanding issue, we believe that it might be feasible to construct a highly scalable hardware-based accelerator.

## 2 Background and Motivation

Before getting into the proposals details, first we will provide a succinct introduction to the main concepts used by CLA and explain how software limitations might make a hardware-specialized implementation of the algorithm attractive. Although the interested reader might need to track the details in the bibliography provided, we hope that it could be possible to grasp the core components, which are surprisingly simple and yet rather elegant.

---

[2] Since the hierarchy is not considered by the algorithm, from now on, we will denote it just as CLA

[3] Although there is a plethora of proposals inspired in cortical structures, most are not inspired in neuroscience facts but in mathematical constructs.

## 2.1 Sparse Distributed Representation (SDR)

Empirical evidence [39][55] suggests that the neural system represents information using *sparse* activity patterns. In this representation [49], in contrast to conventional binary data representation (also coined *localist* [44]), each bit has semantic meaning. In this way, the data representation is highly resilient to a noisy and faulty environment (as the biological one is), i.e. changing a few bits in the representation always produces a value with "similar" meaning to the original. To convert a localist representation (which can be any multidimensional data representation), an encoder has to be used expanding the original data by a large number of bits (of the order of thousands) where only a few can be *active* at a given time (typically ~2%). Note that, as well as resilience, an inherent property is the low power requirements to send information. For example, a 2048-bit SDR stream of data will require up to 280 bits in a localist binary representation. Therefore, the SDR will require, on average, 3.5x (140 vs 40) less bit activations. Another salient property of SDR is the Union property (to store multiple data in the same representation with a low probability of false positive identification) [3], behaving as a space-efficient probabilistic data structure which for example resembles the properties of Bloom filters [9]. Strikingly, the basic principle is simple: the number of combinations of a few elements in a large set is so large[4] that a low number of coincidences allows the identification of a value with a very low probability of error. Biology, through the massive time scale of evolution, has perhaps reached the same conclusion. The basic principle of the theory on which CLA is based is that the cortex also uses SDR representations to store and retrieve information.

## 2.2 Hierarchical Temporal Memory and Cortical Learning Algorithm

Currently CLA focuses on partially replicating the functionality of the cortical micro-columns. Layer I is mainly used for interconnecting different near columns (in the same region). Layers II/III, usually denoted *inference layer*, are supposedly devoted to predicting the state of the column in the next input steps. Layer IV, denoted *sensory layer* [12], handles the input signals to the column coming from the senses (and motor command copies). Layers V and VI, handle the output from the column to sub-cortical brain regions and lower level regions in the hierarchy respectively. Inference layer projects its outputs to higher level regions in the hierarchy. The thalamus acts as a relay point for the inputs (i.e. incoming axons to the region from senses or higher cortex regions) [16]. In order to avoid storage redundancy, the hypothesis of the HTM is that different regions are connected hierarchically (from layer III/VI to layer IV of other cortex regions). Therefore, to identify the pattern of a particular letter in a word, the cortex might be using a single lower-level region. There are multiple layers in the hierarchy (e.g. from simple lines to poetry). The key point is that the same regions can be reused by different regions at the next level and throughout the hierarchy the semantic meaning of the activation patters will be higher.

Unfortunately, the organization of this hierarchy (i.e. how the layers in different regions interact) is not well understood by the neuroscience ([5],[10], [50], [52]). Therefore, the connections to other regions are not actually considered by the CLA algorithm. Although there are ongoing research efforts to support it [56], we have focused our attention over a single region. Even at the current state, CLA algorithm is enough, from a practical standpoint, to produce a useful system. The reader should be aware that the purpose of the theory is not to mimic brain functionality (at least, at this point in time) but just to seek inspiration in the techniques learned through evolution to implement a memory system (e.g., useful in a task that is unsuitable for a von-Neumann architecture).

The CLA defines the term column, which is sufficient to handle hierarchy-less prediction (see Figure 2). A *proximal*[5] *dendrite segment* [25] could be connected to a subset of the bits of the SDR encoded input (which might be provided by a localist-to-SDR encoder). This restriction models the fact that the input axon´s potential (i.e., spikes) will be observable from a subset of the columns. This segment models the dendritic growth of the feed-forward connection of the system. It is well known that dendritic growth is responsible for the learning in the cortex [22]. In contrast with other artificial neural networks (ANNs), each synapse of the segment is characterized by a binary value, i.e. it is connected or not. For a given encoded input, in each proximal dendrite segment, the number of active synapses is determined, i.e. the number of active input bits connected to the segment with a formed synapse (this is called *input overlap*). Once this is known, like in biological systems, an *inhibition process* begins and only the top ~2% columns with most active synapses (e.g. largest overlap) are selected. The remaining columns are inhibited. The synapses to active input in the winning columns are strengthened and synapses to inactive inputs weakened. In order to handle learning, each synapse connection is tracked with a *permanence* value. If the value is above a predefined threshold, the synapse is considered connected. At boot time, the values are chosen randomly near the threshold value. Short integers are sufficient to store the permanence. In the CLA terminology, this is called *spatial pooling*. Therefore, the Spatial Pooler is in charge of producing a stable SDR-compliant representation of each input value (which can be noisy) [34].

When a column is activated, i.e. wins the inhibition process, (*temporal*) cells are in charge of predicting whether the column will be active in the next cycle or not. Each column will have a few tens of cells. Multiple temporal cells per column allow the input value in different contexts to be represented (i.e. the memory is capable of predicting high order sequences). Even with a low number of cells per column, the number of "contexts" that the system can store for the same value is enormous. For example, in a system with 2048 columns and 32 cells per column $40^{32}$ different temporal contexts can be represented for the same value. Each cell might predict the status of the column in the next input in the sequencer. For this, it uses dendrite segments modeling column-to-column relationships *called distal dendrite segments*[6]. Each distal dendrite segment stores potential synapses with temporal cells in other columns in the cortex. The rules for handling these synapses are similar to the proximal segment. If some of the segments of the cell reach a given threshold (number of synapses connected), the cell enters in *predictive state*, which means that this column will be active in the next input value (or epoch). When a column is not predicted correctly, all the cells in the column try to connect to the previously seen sequence (performing a burst). Firstly, by constructing new distal segments on the fly (according to the previous remote activations) and secondly, by looking for cells in the next epoch that predicted the activation. Intuitively, we would use the synapse between different columns in the system using a snake-path through cells in different temporal contexts. The CLA terminology used for this task is *memory sequencing* and it is done in the Temporal Memory [25].

---

[4] For example, $\binom{2048}{40} \approx 10^{84}$, i.e. more than atoms in the Known Universe (~$10^{76}$-$10^{82}$)

[5] Note that although the name comes from the term used for dendrites close to the soma (or cell nucleus) in pyramidal neurons (they are the most numerous excitatory neuron types in mammalian cortical structures), CLA does not require modeling the neurons at low level.

### 2.3 Encoding and Classification

To provide a SDR representation in a practical scenario, an encoder is needed. There are a few rules that an encoder should obey in order to fulfill the SDR properties [3]. For a scalar encoder:

1) The SDR representation of similar scalars should have a high number of set bits in common. Overlap should decrease smoothly as scalars become less similar.
2) The SDR representation of dissimilar scalars should have very low overlap.
3) The SDR representation for a scalar must not change during the lifetime of the system.

Such conditions can fulfilled using a really simple approach (e.g. by constraining the range of values that can be represented) or rather complex (e.g. with large memory requirements and/or large encoding costs). To better understand how SDR works, next we will describe a simple, yet hardware feasible encoding strategy. Let's assume we need to encode $n$-bit, $L$, positive integer into a $k$-bit SDR representation $S$ with $w$ active bits. We use $seed_1 = L \ div \ w$ and $seed_2 = L \ div \ w+1$ as the seed of a pseudo-random generator. Let $R^1$ be the set composed of the initial $w$ unique elements generated by the first seed (where the operation modulo $k$ has been applied). $R^2$ set is generated by the second seed (with indexes not present in $R^1$). We discard the initial $w$-$L$ $mod \ w$ indexes of $R^1$. We set the following $L \ mod \ w$ bits in $R^1$ in $S$. The remaining $w$-$L$ $mod \ w$ bits to be set in $S$ are chosen from $R^2$. Since pseudo-random generators are deterministic, following this approach we can fulfill the three rules previously stated. The probability of having the same two numbers with the same encoding is negligible.

It should be noted that this approach does not require storing the conversion, just a pseudo-random generator and the logic needed to apply the algorithm. Since other non-scalar time series can be remapped to scalars, we can consider that the encoding problem is not a remarkable issue and therefore leave it out of this study.

In contrast with encoding, classification is application dependent. For example, detecting anomalies in a signal is straightforward but predicting multiple steps in the future can be very memory intensive. To achieve the desired flexibility an optimal way is to run the classification problem in a general purpose core, such as [13]. It should be noted that classification is the single non-biologically inspired component in CLA.

### 2.4 Software Limitations and Hardware Opportunities

ML flourishing [4][6][33] has been motivated by the massive raw computational power of state-of-the-art heterogeneous multi-GPU/multi-CPU systems. This has enabled the use of a consolidated theory in increasingly challenging problems, jumping from simple pattern recognition of handwriting [31], to enabling a machine to win in a complex game against the best human [57]. The algorithms beneath these problems are suitable for data level parallelism, where GPGPU model excels [15]. Recent systems, such as the nVIDIA DGX-1, designed mainly for deep learning problems, exceed 2PFlops of computing power in a 42U rack (note that the fastest supercomputer in the world delivers ~33PFlops).

In contrast, CLA's inherent nature makes it difficult to exploit such a paradigm. The synapses, in spite of requiring much simpler computations, can change dynamically. This difficult data level parallelism extraction will make GPGPU quite inefficient. Currently the support for this computing model in NuPIC is not even initiated. Perhaps, as happened with DNN in the past, CLA might not be able to take full advantage of the current and forthcoming hardware advancement, which might constrain the reach of the idea.

A CLA custom hardware accelerator will not only overcome these limitations by breaking the performance/energy barriers imposed by general purpose CPUs, but will also take advantage of CLA's simple computations and low storage requirements. An insight into this advantage is that the core of most Machine Learning approaches is floating-point (matrix) multiplication and CLA only requires low precision integer

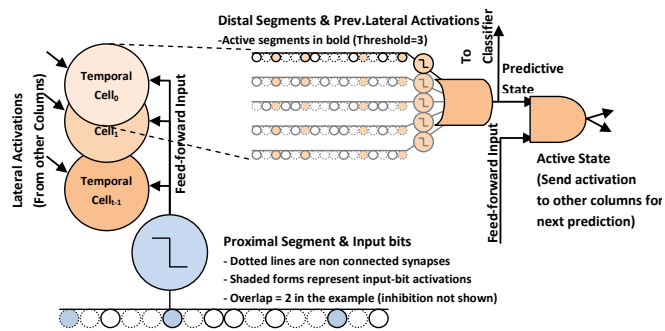

**Figure 2 (a) CLA Column Components, (b) Temporal Memory Prediction.**

---

[6] Although the algorithm used here follows [25], which only includes the basal dendrites, there is an effort to extend this differentiating the role of the apical dendrites to develop a posterior temporal memory.

addition/comparison. For example, according to [27], 32b FP multiplications will require 100 times more energy than to add two 8b integers. Similarly, be able to use only on-chip memory will reduce in two orders of magnitude the memory access energy.

Today ASIC-based DNN, such as [13][14][23][24], is appealing because of the data and precision required fits the resources and the algorithm is well defined upfront. In CLA this might be quite different because of the distinctive properties of CLA (i.e., there is no problem-specific customization, low memory requirements, low precision computing). Then, a hardware implementation of CLA may be more general purpose than a DNN one.

A CLA ASIC might be useful in applications as demanding as data analysis or forthcoming applications such as saccade-based image recognition. For example, processing many millions of streams of data concurrently and using unsupervised learning in order to detect anomalies might be feasible. It is not easy to forecast the potential openings, but if we are able to perform fast CLA Natural Language Processing (NLP), such as [53], it could be possible to tackle challenging problems.

Finally, to really explore the full potential of the hierarchical organization, and propose and validate theories about the underlying and unknown working mechanisms of the cortex, a hardware implementation might be useful. Under these circumstances, it is appealing to explore the feasibility of a silicon-based implementation, as this paper does. Next, we will introduce the architectural details of a potential implementation that we have called CLAASIC.

## 3    About the Feasability of CLAASIC

CLA's basic assumption is that synaptic plasticity (through dendritic growth [34] as a consequence of the back propagation of cellular action potential [22] ) is the key element used by the cortex to learn. The hypothesis is that the information will be stored in the relation between columns, defined dynamically depending on the connections established via on-line learning. Therefore, the storage capacity is proportional to the product of the number of columns by the maximum number of connections per column. The connectivity of the neurons can potentially be very high (the dendritic splines can provide up to tens of thousands of potential synapses). Nevertheless, most of those synapses are not active (i.e., the pre-synaptic axon is too distant from the dendrite) or multiple active synapses correspond to the same pair of neurons (as a redundancy mechanism). Instead of electrically replicating the morphology of biological systems, which perhaps is unattainable, we will embed this functionality in a packet-switched network. We will focus our interest on how to organize and optimize the communication substrate to emulate axon spikes and correctly apply the prediction and learning algorithms of the CLA. Instead of using synapses to establish a connection between two columns, we will use memory structures attached to each router modeling dendritic segments and the required logic performing the spatial pooling and providing temporal memory. Figure 3 (a) presents a high-level description of the proposed architecture. We will assume that the encoder, i.e. the component in charge of converting a localist input into a SDR representation, and the classifier, i.e.

the component that will be in charge of performing the intended purpose of the system (e.g. detecting anomalies in the input sequence, predicting the forthcoming input sequence, etc.), will play the role of I/O interface. We will implement the actual CLA mechanics in a component called the Columnar Core (CC). In this particular example, we will use a sixteen-core system connected by a packet-switched mesh network. Figure 3(b) shows a high-level sketch of a CC. In this case, we will assume that each CC has B columns and $t$ temporal cells per column. Like the biological cortex, the system is homogeneous. Next, we will briefly discuss the requirements of each component to later focus our attention on the most relevant one: the communication substrate.

### 3.1    Communication Requirements

The interconnection network has to handle all the traffic generated by the CLA algorithm. The traffic has four purposes: (1) input traffic incoming through the Encoder, (2) inhibition traffic, (3) lateral activity due to temporal cell activations and (4) column activation and predictions sent to the classifier. This activity will be done at logical level using packets instead of physical wires. For example, each output bit of the Encoder will be connected to a statically defined set of columns. Then, for a given input, each active bit in the SDR representation will be transformed into a multicast packet, addressed to the potentially connected columns. The Encoder will need a table with the relation between columns and inputs. Therefore, a multicast packet will emulate the axon spike. Similarly, when a column enters a predictive state, a unicast packet will be sent to the classifier. Note that biological systems do not have an equivalent classifier and it is necessary, in absence of hierarchy, to make the system practical. The equivalents to the encoder are the senses [39].

Internally the router will receive inputs from the spatial pooling logic (column overlay used in inhibition) and the temporal memory logic (cell activation events). Those inputs should be sent to the potential receptors as packets. The CLA software algorithm assumes that in most cases all the columns in the system should be aware of them, i.e. the potential receptors are all the columns. For example, for global inhibition (which is the default approach), any column should be aware of the input overlap of the rest of the columns. The overlap is computed as the count of connected synapses in the proximal segment of the column for a given input. With this information, the pooling logic is able to determine whether the current column is among the 2% with highest overlay and to feed-forward the temporal memory logic. Similarly, to construct distal segments, although probabilistically limited, the algorithm assumes that each column is aware of all the *(temporal)* cells in predictive state. This is equivalent to assuming that the axon spikes are broadcast to all the cells in the system.

At first sight, the communication requirements are not easy to handle. There is a large amount of multicast/traffic that will require broad network bandwidth and large energy consumption. Additionally, any of the computations performed in the computing layer should be done accessing only local information. In order to scale the system to thousands of CCs,

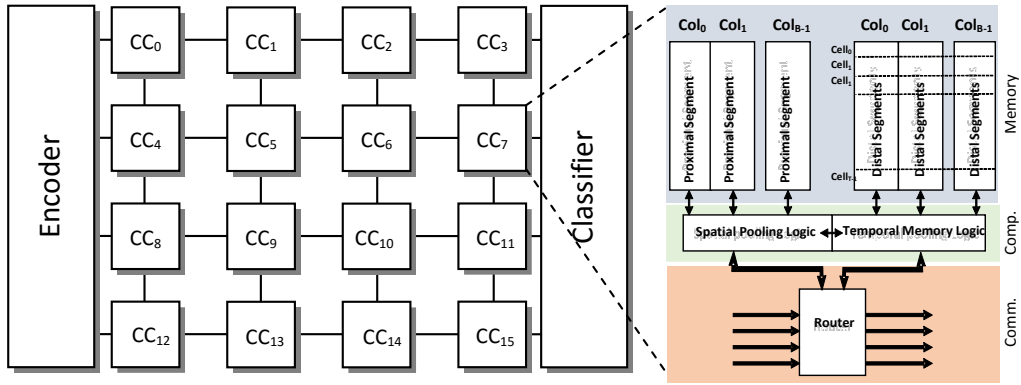

**Figure 3  (a) The Columnar Cortex, (b) High-level description of a Columnar Core (CC)**

we cannot rely on any centralized component. It might be complex to achieve synchronized behavior.

### 3.2    Computing Requirements

There are two stages in the CLA algorithm that have to be applied sequentially, once all the traffic of the current cycle has been drained from the network:

**Spatial Pooler**. The logic in charge will evaluate the input activity. The computing logic will evaluate the input overlap with its proximal segment (i.e. the number of synapses connected to an active input) and broadcast its value (assuming global inhibition) to the rest of columns in the system.

Inhibition logic might be pretty straightforward (assuming global inhibition). Therefore, in each column we only need to compare the remote overlap of the incoming packet with the current one. If the number of columns with higher overlap is above the activation limit, i.e. ~2%, the column is self-inhibited. To break ties, additionally to the input overlap, each column will include its ID in the inhibition packet.

The synapses in the proximal segment table of the active inputs will be adapted if the column wins the inhibition. Therefore, the spatial logic will require a comparator, a 4-bit adder and a counter, the maximum overlap required being ~Log₂Input bits. For a 2048 input encoder, 12 bits will suffice. Note that although column boosting might be required to achieve a balanced column activation pattern [18], according to [34], it is not cost effective to do so in resource-constrained environments like an ASIC.

**Temporal Memory**. The logic should evaluate any lateral activity. If we assume that the axon of the *temporal* cells is global, a broadcast will be generated. The incoming spikes will include the original column and original temporal cell. These will be kept in a list of current activations. Once the current epoch is complete, the logic will determine for each column whether the activation was correctly predicted. In this case, the corresponding distal segment of the temporal cell in predictive status will be updated accordingly (i.e. performing dendrite growth). If the column wasn't correctly predicted, the logic should keep the activations from the previous cycle to search for the closest distal segment (or create a new one if there is none). From a naive perspective, this can be difficult since it requires an extensive search through all the dendritic segments of the column.

The temporal memory should determine for the current activations which dendritic segments in the column are active. The *(temporal)* cells with an active dendritic segment will generate a broadcast/multicast to the network. Finally, the columns that were not predicted correctly will produce a burst, like a biological system does [28], which basically is equivalent to putting all the temporal cells in the column in active state (but selecting only one for doing the learning).

### 3.3    Memory Requirements

The precision required by the algorithm is extremely low. In a practical problem such as [30] from full 64b FP precision to 4-bit integers, there is no appreciable performance loss (<1%). The reason for this is that there is low sensitivity to learning rates. In this application Temporal Memory uses steps of 0.1 (with permanence between 0 and 1). The spatial pooler uses smaller values (0.08 for learning and 0.003 for forgetting), which can be modeled with 8-16 levels (using probabilistic subtraction and addition during the learning). This is biologically plausible, since cortex dendrite growing/shrinking is a stochastic process [47].

The proximal segments will store the permanence of the synapses with each potentially connected input bit. Note that each bit of the SDR representation produced by the encoder is potentially connected (i.e. a synapse might be formed) to the chosen subset of columns at boot time. In general, we can assume that each bit can be connected to any column in the system. Therefore, the proximal segment has to have one entry for each potential input. In practice, each column will be connected (i.e. a synapse will be formed) to a very small subset of encoder inputs. Therefore, we might structure the proximal segment as a conventional cache indexed by the input index. In practice, having capacity for 64-128 entries in a 2K column system seems to be enough.  The permanence value must be stored there. Reduced precision weights in DNN have a much more adverse effects in system performance [23][24]. For example, if we assume a 2K column system with 1K inputs, the aggregation of all cortex proximal segments will require (including tags) between 0.25MB and 0.5MB. However, issues such as conflicts have to be considered. The inherent nature of the SDR representation suggests uniform usage.

In a naïve approach, each distal segment will require as many synapses as columns in the system and each temporal cell might have an unbounded number of segments. In practice, for a 2K columns system, having 128 segments with 40 synapses per cell provides similar results to an unbounded system [25]. Therefore, excluding tags, ~80KB will be required for 4-bit precision per column, i.e. 160MB for the whole system. This amount seems feasible to achieve an on-chip SRAM memory storage such as is presented. Although orthogonal to this work, note that this is a raw amount that can be greatly reduced using the proper techniques. Exploiting CLA's noteworthy fault resilience (>50% rate of faulty cells can be tolerated [25]), it could be possible to reduce the final storage requirements significantly.

### 3.4 About the Temporal Cost of the Computing Phase

A key insight is that the learning (the most complex part of the algorithm) is outside the critical path. Since prediction only requires comparisons and counter increments, it is reasonable to assume that the time required to perform the prediction will be memory bounded. Since the memory required per column is fairly small, with the proper SRAM configuration, it could be possible to perform this operation in one cycle (on spike arrival).

The learning algorithm, especially distal segment formation, is more complex. Nevertheless, given the random nature of column activation, we can assume that the learning on one column will be done with sparse frequency, i.e. only ~2% of the time. Therefore, the time budget for the learning is 50 times larger than the prediction. Although the logic in charge of this is not analyzed in this work, it is reasonable to consider it not relevant from the hardware perspective (both in time and area).

## 4 Communication in the Columnar Cortex

We have identified three major problems in the CC: communication and synchronization, temporal memory logic complexity and distal segment organization. From a scalability standpoint, the most relevant seems to be the former one, since the necessary scalability appears to be a key element in the cortex.

In biological systems, the critical difference between species seems to be dominated by the number of neurons and not the number of synapses per neuron. For example, the mouse cortex [45] and the human cortex [38] has roughly the same number of active synapses per neuron (~$7 \cdot 10^3$ synapses/neuron and ~$7.2 \cdot 10^8$ synapses/mm$^3$). The difference is the size of the cortex which is much larger in the human case (~$112$ mm$^3$ versus ~$650$cm$^3$). The biological facts suggest that somehow, the inner-CC issues are not a significant problem since the tables, and the time required by the temporal memory logic will not need to scale up with the total number of columns.

In contrast to the mouse brain, the number of "inactive" synapses is much larger in the human brain, since the potential connection will be proportional to the total volume. Clearly, when we increase the number of columns in the CLA algorithm, the demands in order to communicate and synchronize different CCs will be substantially higher.

Next, we will discuss the key elements of the communication substrate.

### 4.1 Network Characteristics

Since all the spikes will be modeled as multicast packets, to obtain a reasonable performance, the router requires multicast support (i.e, in-network replication). This can be done effectively with little to no cost impact using a router such as [29].This approach will also reduce energy requirements, since the copying of the packet is performed near the destination, and it will achieve a low latency, since there in no injection serialization.

The packet size required is rather small. Inhibition traffic will require overlap and tie-breaker ID ($Log_2NumColumns$ +$Log_2NumEncoderActiveInputs$). Lateral activity will require the source column and temporal cell ID ($Log_2NumColumns$+$Log_2NumTemporalCells$). Input activity will require source ID ($Log_2NumInputs$). For a 2048 column/input system, with 32 temporal cells per column, the size required will be 22, 16 and 11 bits respectively. Although, these sizes are much smaller than in a conventional CMP (where in most cases, the packets are around tens of bytes), the cortex organization or further enhancement (such as Section 4.5) might require adjusting the bandwidth availability (i.e., link width).

Since the individual latency of a packet is not critical, a low-degree network with narrow links might satisfy the requirements. High-degree networks will require increasing the complexity of the routers and the wiring cost. Therefore, 2-D Torus or Mesh [20] might meet these requirements. Although not explored here, as with biological neurons, CLA gracefully tolerates a faulty/noisy input [25][17]. Therefore, it will also tolerate a faulty network. With a fault tolerant network such as [42], it could be possible to scale up the system size without yield issues even using wafer-to-wafer 3D integration under aggressive technological nodes.

### 4.2 Synchronization

Looking at the algorithm, there are four main phases: computing overlap of the proximal dendrite with the current encoded input, determining the winning columns in the cortex, determining the lateral activity in each temporal cell in the column and producing the prediction. Overlapped with those phases, the adaptation (i.e. learning) of the synaptic segments is performed.

The main difficulty of running those phases in a fully distributed way is to know *when* each one should be done. For example, input overlap should not be run until all the input activity is received (i.e. each column is aware of all input spikes). Since there is no acknowledgment message of axon spike reception, each CC should be aware when to run the corresponding part of the algorithm. Similarly, inhibition cannot be activated until each column is aware whether it is within the most active ones and finally, prediction cannot be done until the lateral activity of the related temporal cells is known. The simplest, yet most efficient way to circumvent this problem is to drain the network content before progressing to the next phase. If the network is empty, there is a guarantee that all the influencing packets will already have arrived at the destination.

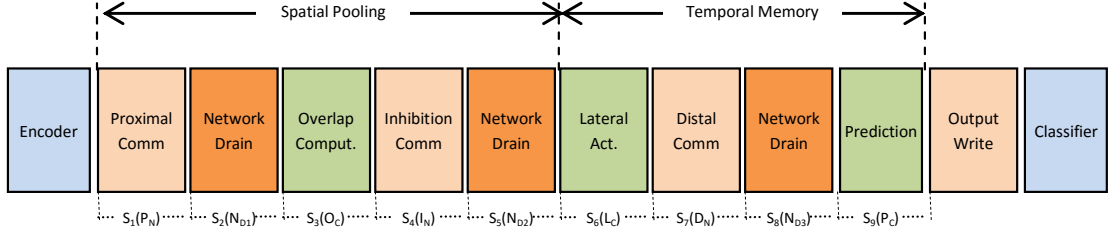

**Figure 4 Stages in CLA Algorithm**

Figure 4 details all the stages required for the CLA algorithm. Besides encoding and classifying, there are nine stages, three of them perform computation in the spatial and temporal logic (S3, S6 and S9), three correspond to the axon spikes (S1, S4 and S7) and another three are required to drain the network (S2, S5 and S8).

The problem of synchronization is then reduced to providing a scalable network drain mechanism. To guarantee the scalability of such a mechanism, we need a cost-effcient way to do so within the network itself. A simple approach is to use dimensional order routing (DOR) [20] and inject a special broadcast packet, denoted *broom* packet, into the extreme Columnar Cores from the smallest and biggest ID (in the example in Figure 3, these should be $CC_0$ and $CC_{15}$ ). These packets will be allowed to progress to the next routers only if the local router has no more packets in the injection queue and the transit buffers at the ports where the router has received the copies of the packet are empty. The packet is replicated through the remaining ports. For example, when $CC_5$ receives the $CC_0$ broom packet from $CC_4$ and $CC_1$, we know that there are no normal spike packets that might affect the columns handled by $CC_5$. When the transit queues from W and N are empty, the router replicates the $CC_0$ broom packet through the S and E ports. This operation will be applied in the whole cortex until the

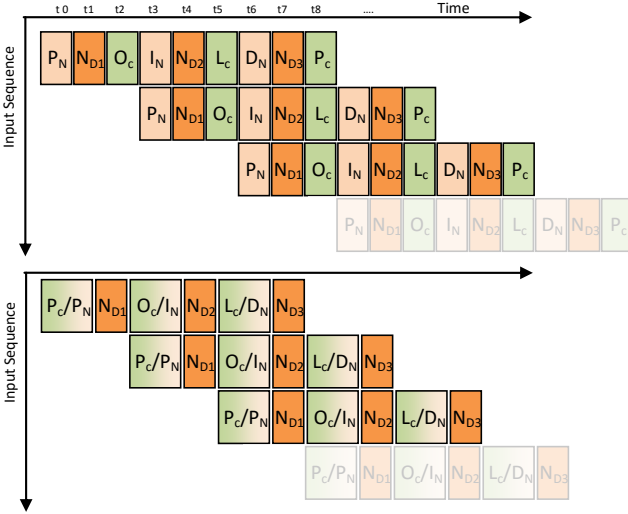

**Figure 5 (a) CLA pipelined algorithm, (b) Overlapping communication and computation**

columnar core $CC_{15}$ receives the broom packet from $CC_0$. At this point, $CC_{15}$ is aware that there are no packets in the network and it can progress to the next stage in the algorithm. Similarly, when an intermediate CC receives all the broom packets from $CC_0$ and $CC_{15}$, it knows that there are no pending packets in the network for it. It should be remarked that this mechanism operates in a fully distributed way and will scale according to the network's available bandwidth.

We hypothesize that in biological systems, this drain is not required because the input rate of change is slow enough to guarantee that the spatial and temporal memory are handled satisfactorily. When the input rate is too high, the system will be unable to learn or predict. As a naïve example, an excessively fast image rate of change will be perceived by the visual cortex as noise. Although a similar solution can be applied in our case, we think that encoder and data are not evolutionarily tuned like in biological systems and perhaps will require an excessively long worst-case delay to work correctly in corner cases.

### 4.3 Pipelined Algorithm: Communication and Computation Overlap

The nine stages in the algorithm will certainly require a substantial amount of time and energy. In particular, the network seems to play a fundamental role, since it is foreseeable that the time required to propagate the axon spikes will be large. Nevertheless, if we look at Figure 4, we can identify stages like in a general purpose processor.

Therefore, we can use the same optimization techniques used there. In particular, we can pipeline the algorithm reducing the stages per input sample to three. Figure 5.(a) shows how that organization will be beneficial once the pipeline is loaded. The idea is to start computing the overlap of the next input in the sequence as soon as we know the current overlap. Then at $t3$[7], two operations are being performed simultaneously in the network. If we move forward in time we can see how we can overlap three different input operations in a single stage. At $t6$ we are communicating the distal activity associated to the first input value, the inhibition traffic of the second input value and performing the proximal activity of the third datum. At $t8$ we are simultaneously performing the prediction for the first epoch, the lateral activity computation for the second one and the overlay computation for the last one. Even more importantly, we will need only a network drain per input value. Once the pipeline is loaded, we need only three epochs in the input sequence to produce a prediction.

---

[7] This could be a system-status dependent number of clock cycles depending on the computing logic and the network characteristics.

This approach opens up the opportunity for further improvements. We do not need to finish the computation phases before starting to send the outcome of each one (i.e. we can fully overlap the computation and communication phases). As soon as spatial and temporal logic starts to generate axon spikes they can be injected into the network, as Figure 5.(b) shows. Therefore, the number of clock cycles required to process a value in the input sequence will be determined by the slowest portion: communication or computation. The number of cycles required by the slowest one and the clock cycle will determine the time required to process one sample in the input sequence. Finally, network drain should be synchronized across epochs: broom packets are forwarded in the CC router both if there are no packets in the injection queue and transit buffers and if all the local columns have finalized the current epoch (in the spatial and temporal logic). Therefore, network drain operates as a synchronization barrier. Given the discussion provided in section 3.4, if the number of cycles required to perform the communication is larger than the number of distal and proximal packets received per CC, the computation/communication will be fully overlapped. Note that the number of distal and proximal packets received is fairly small (~2% of the columns in the cortex ~2% of the inputs will generate, e.g. for a 2K system each column will require, on average, ~80 accesses to the memory in order to perform a prediction).

#### 4.4 Traffic Aggregation: Colaescing Injectors

Combining multiple columns in a single CC is an interesting approach, from the latency standpoint. To use links between routers with a very short distance can unnecessarily increase the average latency in the network. To optimize this delay, the size of the CC (i.e. number of columns) should be tuned to match the propagation delay with the network clock cycle. This is well known for a Non-Uniform Cache architecture [37]. With this approach, it could be possible to aggregate multiple spikes coming from columns in the same CC in a single packet. Although this might increase the number of flits of the packet, it will reduce the network load.

Finally, the pipelined algorithm opens up the opportunity for additional traffic aggregation. In the case of using global inhibition and distal activity, we can combine actions coming from different stages in the algorithm in a single packet. For example, inhibition can be combined with the lateral activations of the previous epoch.

In order to do so, we assume the existence of coalescent injection queues (similar to the structure used to support non-blocking caches in a von-Neumann processor). Before queuing new packets in the injection buffer of the router, the packets waiting to be injected are checked. If there is a match in the destination mask, the previous packet is modified appending the information of the new one and then discarding it.

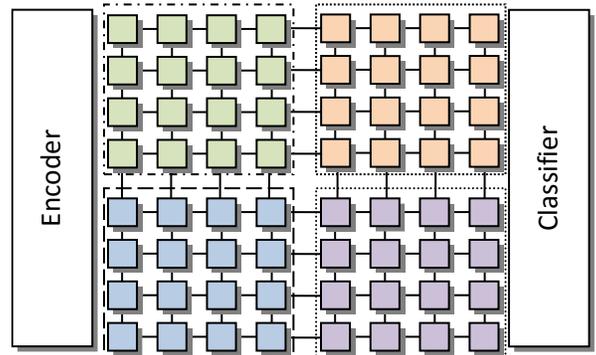

**Figure 6 Scale-out zones Example**

#### 4.5 Scaling Traffic: Scale-out Zones

Biological systems indicate that the best approach to increase the system storage is to increase the number of columns and not the number of temporal cells (and distal segments) per column. As discussed previously, a key point is that the number of synapses per volume is really similar in the mouse and the human brains. The critical difference between the two is the cortex volume, i.e. the number of columns. From a practical perspective, if we increase the number of columns we might reduce the number of distal segments required per temporal cell. Although from a software perspective this does not seem interesting, from the hardware stand point it is really relevant because it might reduce the interconnection cost and perhaps the complexity of the CC. Therefore, in a hypothetical silicon implementation it will be desirable to increase the number of columns as much as the technology allows, i.e. depending on the yield and/or power envelope. Unfortunately, the communication system, as described at this point, might scale up to a limited number.

Distal and inhibition traffic are assumed to be global by the CLA (although inhibition might be local, it is rarely used because the performance falls significantly[8] and a loss of accuracy is incurred [34]). From the network perspective, the delay and power requirements will be increased significantly as we increase the number of CCs. Note, that the number of columns involved in the inhibition process is substantially higher than the number of inputs active in the encoder.

Biological systems do not use global communication in such processes. Inhibition, which is performed by inhibitory interneurons [11], should have a localized and static radius. Similarly, distal activity is constrained to the shape of distal dendrites and axons of pyramidal neurons [32]. Proximal traffic is less demanding because the CLA algorithm assumes that potential proximal synapses are limited. At boot time, each column can potentially be connected to a subset of inputs, called the receptive field [40] of the column. Usually the receptive field of each column is a subset of the input bits following a topological arrangement. This improves the accuracy of the system. Coincidentally, it reduces proximal traffic relevance, since the destinations in the multicast packet will be localized in the same region of the cortex.

To circumvent the global communication problem, we propose a simple approach that is based on splitting the network into separate zones and restricting the inhibition and distal traffic within them. We denote these regions as *scale-out zones*. For example, Figure 6 shows those zones in order to increase the number of CCs from 16 to 64. Instead of requiring broadcasts, the traffic generated by columns in any of these zones will be restricted to them. If we need to further increase the number of columns, we can simply increase the number of zones. With this simple approach, traffic will be kept constant.

The encoder, i.e. proximal traffic, selects the potentially connected columns without making distinctions between zones, i.e. the receptive fields are kept constant. The approach we propose is to use as many consecutive values in the encoded input sequence as the number of scale-out zones. In our example, we will use 4 encoders to simultaneously encode four different epochs from the input sequence. Thus, we not only increase the throughput of the system but also the load on each individual column, since a whole representation is scattered throughout the whole system. Additionally, increasing the number of zones will keep the total proximal traffic constant (since receptive field size is kept constant).

In an *n-zone* system, each column only sees an *n-th* part of the input data. Therefore each CC will require an *n-th* part of memory requirements, improving the system scalability. Access times are faster and the time available to accommodate the computation (during the communication phase) is n-times greater, which might allow slower but denser memory technology to be used.

# 5    Evaluation Methodology

## 5.1    Tools and Benchmarks

We have developed an integrated simulator, CortexSim[41], which emulates all the previously depicted mechanisms. CortexSim is influenced and verified against the NuPIC white paper implementation (but columns boost) of the CLA algorithm using the Numenta Anomaly Benchmark (NAB) [30]. The simulator is connected to a network simulator, Topaz [1], in order to obtain precise network timing results and DSENT[48] and CACTI[37] to estimate area and energy requirements. The data sets used in this evaluation are both synthetic and real. We use synthetic data to simplify architectural comparisons and realistic data to provide a notion of the benefits of the accelerator in a real application.

The real data is provided by NAB. The NAB corpus of 58 time series data files, composed by ~350,000 samples, is designed to provide data for research in streaming anomaly detection algorithms. It is comprised of both real-world and artificial time series data containing labeled anomalous periods of behavior. The majority of the data is real-world from a variety of sources such as Amazon Web Services metrics, Twitter volume, advertisement clicking metrics, traffic data, and more. The data includes anomalies that are annotated by human reviewers, following a strict procedure. This data is processed using many anomaly detection mechanisms, and serves to compare with

CLA in this particular task. Each data set has a probationary period (~10%), during which the detector anomaly detections are ignored. Note that in each time series the detector its reset. Although the data diversity is really high, the parameters of the cortex are constant in all experiments. Under these conditions, NuPIC is able to reach 65% successful anomaly detection whereas other state-of-the-art approaches are 20% behind.

For the synthetic workload, we will use periodic series of 32-bit integer data generated from randomly defined polynomials (up to fourth degree with randomly chosen coefficients). A limited number of points from each one are defined for twenty values of x, defining a temporal set. We will repeat each temporal set until it is learned by the system. We consider that the time series is learned when the number of elements in the sequence with no miss predictions (i.e. no column bursts) is equal to half of all the data points. The rationale of this is to keep half of the time for learning new sequences and half of the time for predicting them. Therefore, half of the epochs will produce the extra traffic of column bursting or low overlap inhibition that a new input sequence appearance will generate. The second half of the time, the system will have a stable representation of the input, being less demanding for the network. The number of temporal series (i.e. polynomials) used to fulfill strict 98% confidence intervals is around ~50.

In both cases, the classifier we will use is an anomaly score estimator. This is the simplest one and just provides the fraction of the miss-predicted columns.

## 5.2    System Configuration

In regard to the CLA configuration, we mimic the one used by [30]: 45x45 column cortex with 32 temporal cells per column (with up to 128 distal segments), with global inhibition, a 2045-bit SDR encoder with a diameter in the receptive field of 32. In contrast to [30], we use a new SDR encoder, succinctly introduced in section 1.3. This encoder is simple to implement in hardware (only requiring a pseudo-random generator and some logic to build the SDR representation) this encoder improves NAB detection rate by 1-2% compared to the one employed in [30], denoted Random Distributed Scalar Encoder (RDSE). The encoder, in synthetic workload, has full integer precision. In NAB we limit it up to 130 levels of quantification (as RDSE does in [30]). All CLA parameters are kept constant throughout all the evaluation.

In regard to the network, we employ a 2D Torus topology with a conventional router with deterministic DOR, using bubble flow control [43] as a deadlock avoidance mechanism (single buffer of 160 bytes per port, no virtual channels), 4-cycle pipeline, and using virtual-cut through flow control [20]. We assume that the link wires use low-swing links and require a clock cycle to travel from router to router. The clock cycle, conservatively, is 1 ns. We use dimension-order replication for multicast deadlock avoidance. The router has the embedded network drain mechanism depicted in Section 4.2.

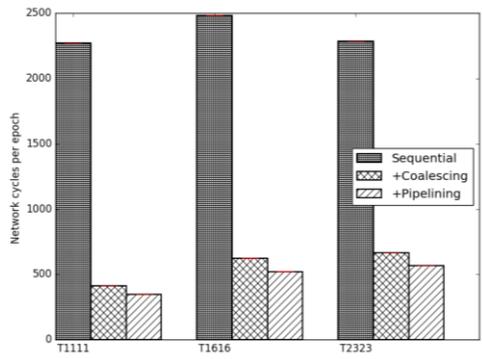

**Figure 7 Network clock cycles per input epoch for different 2-D square mesh sizes with different algorithm optimizations (16-byte links)**

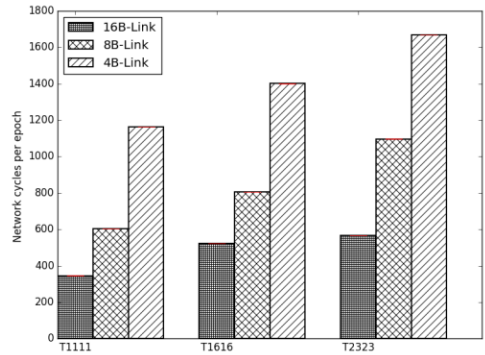

**Figure 8 Number of clock cycles per input epoch for different 2-D square mesh sizes using pipelining and coalescing injectors with different link widths**

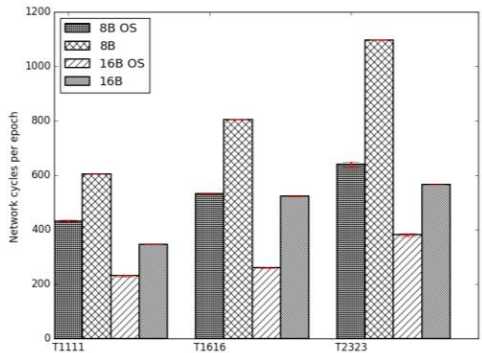

**Figure 9 Network clock cycles required to process an epoch with 4 out-scaling zones compared with no out-scaling with 8 to 16-byte links**

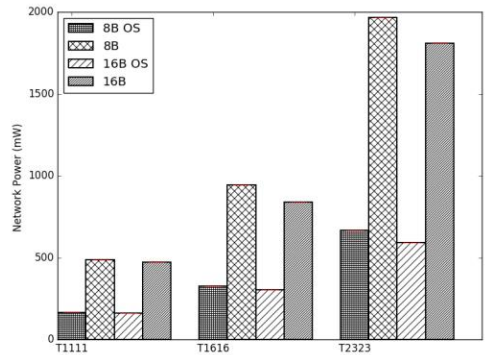

**Figure 10 Power required processing an epoch with four out-scaling zones compared with no out-scaling with 8 to 16-byte links**

## 6    Performance Results

### 6.1    Synthetic Benchmark

Figure 7 shows the number of clock cycles required for each input epoch, for 11x11, 16x16 and 23x23 tori. (i.e. different numbers of columns per CC). As can be appreciated, for a plain approach (sequential) there is little effect on the network size, i.e. network contention dominates. This is due to the high load that the network supports. Adding more nodes increases the raw bandwidth, which in the case of 23x23, allows the time required to process an input to be reduced slightly. When we add coalescing injectors, the delay is reduced four times, because the inhibition traffic is not able to use the links efficiently. The contention reduction allows this improvement. Adding pipelining opens up the opportunity for further traffic reduction, although limited packet size (80bytes) allows limited aggregation. Nevertheless, the true advantage of pipelining is that computation can be fully overlapped with communication, i.e. we might actually need only around 500 cycles to fully process an input epoch. Instead of more advanced routers, such as [2], which has support for adaptive multicast and routing, with a lower implementation cost, we decided to use a canonical router in order to clarify the effects of each change.

Since the number of accesses per column is approximately 80 (40 for proximal traffic and 40 for distal traffic), computing might need ~1300 cycles in 11x11,~640 in 16x16, ~300 cycles in 23x23. It seems that the most suitable network for this configuration is the 16x16 system.

The reduction in contention of these techniques opens up the opportunity of reducing the cost of the network by narrowing the link widths. Figure 8 provides the results of this change.

To improve these figures, the out-scaling zones might be useful. Figure 9 suggests that moving from 1 zone to 4 zones, we can significantly reduce the communication cost. It could be possible to process an epoch in ~200 network cycles. This is because the inhibition and distal traffic only has to reach a fourth of the network. Each column on average will perceive one fourth of the remote spikes, therefore the CC will need a fourth of the memory accesses in order to perform the prediction. Consequently, it seems feasible, using a 16x16 system, to achieve ~160cycles.

Figure 10 shows the total power (both active, and leakage) required by the network. In the out-scaled configurations the power is ~250mW for 16x16.

### 6.2    Realistic Benchmarks

Figure 11 to Figure 14 show how the accelerator will perform using it in the anomaly detection problem. The corpus of the benchmark is composed by different families of data, grouped and tagged on the x-axis. The error bars represent the variability of the performance metric within each family. According to Figure 8, a 16x16 with 16B-wide links is the most interesting. This configuration is also the best performer when four out-scaling zones are used. Under this configuration the latency to process each data set is ~500 cycles and just ~300 when out-scaling is used. This means that the 350,000 data of the whole benchmark can be ingested by the accelerator in just 0.175-0.1 seconds. The average power required by the whole system, under these working conditions, will be between 1.2Watts and 350mWatts. This is equivalent to other DNN accelerators, such as [14][24].

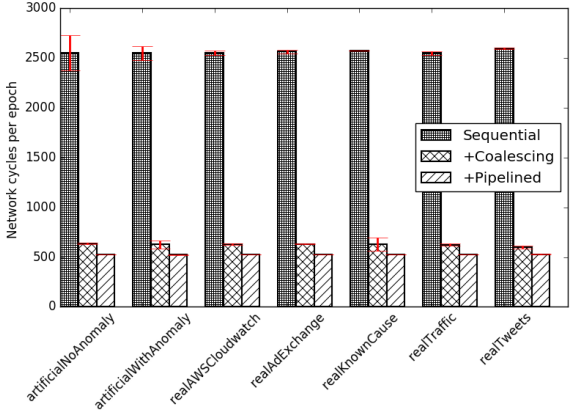

**Figure 11 Network clock cycles per input epoch with each NAB data set for a 16x16 CC Configuration with 16-byte wide links, varying optimizations**

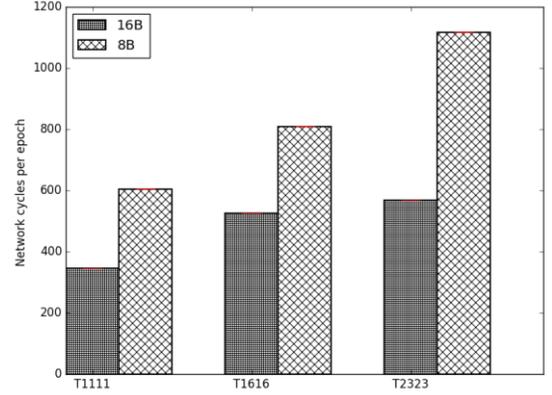

**Figure 12 Average Network clock cycles for NAB with different system sizes and network bandwidths**

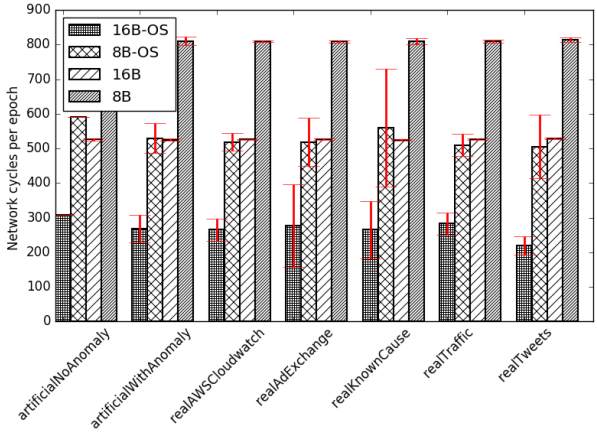

**Figure 13 Network clock cycles per input epoch with each NAB data set for a 16x16 CC configuration with and without out-scaling zones**

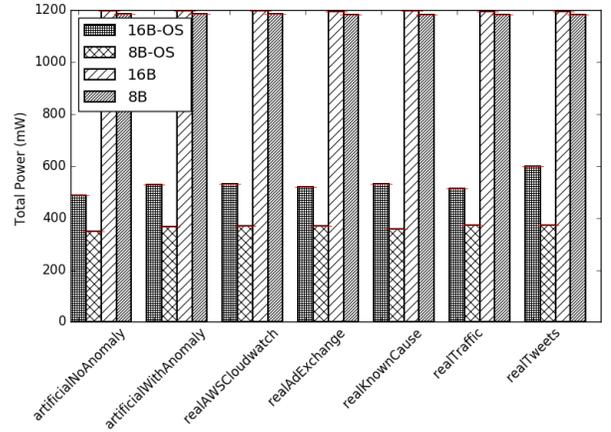

**Figure 14 Full chip power requirements with each NAB data set for a 16x16 CC configuration with and without out-scaling zones**

The latest NuPIC implementation in a 2-socket server based on Intel Xeon E5640 running at 2.4Ghz with 60GB of memory requires ~3000 seconds to run the detection phase in a single core and requires approximately 170Watts. Running in 24 threads, it takes 234 seconds and 450Watts[9]. If we take into account the time and power consumed by the accelerator, CLAASIC is between $3 \cdot 10^4$ and $1.8 \cdot 10^4$ times faster. In terms of energy, the efficiency with respect to a single core is between $1.5 \cdot 10^8$ for the single thread execution (versus the more efficient configuration) and $3.4 \cdot 10^6$ for the 24-thread one (versus the less efficient configuration).

Out-scale improvement in speed and efficiency comes at the cost of accuracy. In practice, the average anomaly detection rate falls 9% compared to the standard configuration. Note that none of the remaining parameters of the CLA are changed when out-scaling is used.

### 6.3 Approximate Power and Area costs

To compare CLAASIC with other deep-learning accelerators, first-order cost figures have been provided. Although the memory implementation has not been detailed in this work, a rough approximation using a CACTI model for a 256-bank 160MB SRAM (power results have been used in the previous section) was carried out. According to the CACTI and DSENT models, a 16x16 system will require ~43mm² (0.154mm² per memory and 0.014 per router). Therefore, it seems feasible to scale up the system size (i.e. number of columns) without much trouble.

In this analysis, learning cost has not been considered, because the time available to perform this task is substantial (for 16-byte links and a 16x16 network, it is ~23K[10] cycles without out-scaling and ~35Kcycles with 4 out-scale zones). Therefore, this

---

procedure could be optimized greatly, to the point of making the cost negligible.

Both energy and power cost can be considered as a worst-case value. On the one hand, the most memory intensive snoops (which correspond to the distal traffic) can be filtered out. On the other hand, CLA fault resilience [25] could be taken into account to minimize the total memory required. In contrast with other approaches, such as [24], we did not need to reduce the number of synapses compared to the software counterpart.

# 7 Conclusions & Future Direction

This exploratory journey provides a suitable design proposal for a cortex-inspired hardware accelerator. A priori, the solutions presented enable the biggest challenge to be dealt with: the communication substrate. From the evidence gathered, from an engineering standpoint, this is not an issue.

The next steps should address the implementation of learning logic and the dendrite segments. Additionally, the use of other practical problems for the CLA as well as anomaly detection will be considered. In particular value prediction might be interesting. For example, combining CLAASIC with a conventional von-Neumann core; we could carry out the classifier task in the regular core, while the CLA algorithm can be executed in the accelerator with a much higher efficiency and speed.

Since CLAASIC multichip organizations are feasible, the proposal is amenable for a hypothetical hierarchical organization. Note that inter-region connectivity will be much sparse [46], therefore seems practical to fit such communication requirements within the constrained bandwidth of I/O. The fault tolerance resilience of CLA and low energy requirements, suggest that to use emerging technologies, such as 3D stacking and non-volatile memories, will be achievable. Consequently, in our view, HTM/CLA might reach biological-level raw capabilities using hardware such as the proposed in this work.